\documentclass{JHEP3}
\usepackage{graphics}  % needed for figures
\usepackage{epsfig}
\usepackage{amssymb}   % for math

\IfFileExists{srcltx.sty}{\usepackage[active]{srcltx}}

\newcommand{\be}{\begin{equation}}
\newcommand{\ee}{\end{equation}}
\newcommand{\bea}{\begin{eqnarray}}
\newcommand{\eea}{\end{eqnarray}}

\def \eV{\: {\rm eV}}

\def\GeV{\: {\rm GeV}}

\def \cm{\: {\rm cm}}
\def \mt{\: {\rm m}}

\newcommand{\eq}[1]{Eq.~(\ref{#1})}

\def \MN{M_{_{\rm N}}}
\def \tN{\tau_{_{\rm N}}}
\def \fN{f_{_{\rm N}}}
\def \fiN{f_{_{\rm i,N}}}
\def \foneN{f_{_{\rm 1,N}}}
\def \ftwoN{f_{_{\rm 2,N}}}

\title{Collider signatures of sterile neutrinos in models with a gauge-singlet Higgs}
\author{Ian M. Shoemaker \\ Department of Physics and Astronomy, University of California, Los Angeles, CA 90095-1547, USA}
\author{Kalliopi Petraki \\ School of Physics, The University of Melbourne, Victoria 3010, Australia }
\author{Alexander Kusenko \\ Department of Physics and Astronomy, University of California, Los Angeles, CA 90095-1547, USA \\ Institute for the Physics and Mathematics of the Universe, University of Tokyo, Kashiwa, Chiba 277-8568, Japan}

\date{\today}

\abstract{
Sterile neutrinos have been invoked to explain the observed neutrino masses, but they can also have significant implications for cosmology and accelerator experiments. We explore the collider signatures of a simple extension of the Standard Model, where sterile neutrinos acquire their mass after electroweak symmetry breaking, via their coupling to a real singlet Higgs. In this model, heavy sterile neutrinos can be produced in accelerators from decays of the Higgs bosons. Their own decay can yield distinct signals, suggesting both the presence of an extended Higgs sector and the coupling of the singlet fermions to the latter. In the same scenario, a relic matter abundance arises from the decay of the singlet Higgs into weakly coupled keV sterile neutrinos.
The coupling of the Higgs doublet to particles outside the Standard Model relaxes the current experimental bounds on its mass.}

\maketitle
\begin{document}
%%%
\section{Introduction}
%%%%

The mere existence of neutrino masses suggests that additional gauge singlet fermions, the sterile neutrinos, may exist in nature~\cite{Minkowski:1977sc,Yanagida:1979as,Glashow:1979nm,GellMann:1980vs,Mohapatra:1979ia}. The masses of these new particles can range from well below the electroweak scale to close to the Planck scale, and, in the case of split seesaw~\cite{Kusenko:2010ik}, the coexistence of such very different scales is perfectly natural.
If sterile neutrinos do exist, they can have a number of astrophysical and cosmological implications~\cite{Kusenko:2009up}.  For example, a sterile neutrino with mass of a few keV may serve as warm or cold dark matter, depending on the production mechanism~\cite{Dodelson:1993je,Shi:1998km,Shaposhnikov:2006xi,Kusenko:2006rh,Petraki:2007gq,Asaka:2006nq,Laine:2008pg,Boyanovsky:2007ba,Wu:2009yr}.
Production of the same particle in a supernova core can explain the observed pulsar velocities and can alter energy transport in a supernova~\cite{Kusenko:1997sp,Kusenko:1998bk,Fuller:2003gy,Kusenko:2004mm,Fryer:2005sz,Hidaka:2007se,Fuller:2009zz}. The X-rays emitted from decays sterile neutrinos can affect the formation of the first stars~\cite{Biermann:2006bu,Mapelli:2006ej,Ripamonti:2006gr,Stasielak:2006br}. Although the prospects for direct detection of keV sterile neutrino dark matter are not encouraging~\cite{Finocchiaro:1992hy,Gorbunov:2007ak,Ando:2010ye}, there are tantalizing hints of relic sterile neutrinos with either a 5~keV mass~\cite{Loewenstein:2009cm} or 17~keV mass~\cite{Prokhorov:2010us} coming from direct or indirect observation of their radiative decays.
Some models suggest that the Majorana mass of the keV sterile neutrinos may originate from the Higgs mechanism at the electroweak (EW) scale~\cite{Kusenko:2006rh,Petraki:2007gq}. This, and the natural  democracy of scales realized in the framework of split seesaw model~\cite{Kusenko:2010ik}, suggest the possible existence of heavy sterile neutrinos~\cite{Graesser:2007yj,Graesser:2007pc} coupled to a non-minimal Higgs sector.  Here we explore the collider phenomenology of heavy sterile neutrinos coupled to a gauge-singlet Higgs field, and discuss relevant constraints.

In contrast with the other fermions of the Standard Model (SM), sterile neutrinos can have Majorana masses.  Since the Majorana mass term is not forbidden by any local symmetry, it is often assumed that the natural value of the Majorana mass is very large~\cite{Yanagida:1979as,Glashow:1979nm,GellMann:1980vs,Mohapatra:1979ia}.  However, it was recently pointed out that in models with extra dimensions, a broad variety of scales are equally natural, and the Majorana mass can be much smaller than the electroweak scale~\cite{Kusenko:2010ik}. In this class of models, there are no small parameters, but the effective four-dimensional Yukawa coupling and the effective four-dimensional Majorana mass are both exponentially suppressed in such a way that the successful seesaw relation is preserved~\cite{Kusenko:2010ik}.   
One particuarly appealing choice of parameters is realised in the split seesaw model, which uses two heavy Majorana masses to explain the neutrino mass spectrum and leptogenesis, while the third, much lighter right-handed neutrino plays the role of dark matter~\cite{Kusenko:2010ik}.  If the fundamental Majorana mass is below the electroweak scale, there is a possibility of an additional contribution coming from the Higgs sector.  The possibility of a gauge-singlet boson present in the Higgs sector is widely considered as a well-motivated scenario for physics beyond the Standard Model.  In view of the possibly small (and natural) Majorana mass, the coupling of such a singlet to right-handed neutrinos is inretesting both for the collider physics and for cosmology~\cite{Kusenko:2006rh,Petraki:2007gq}.  As in Refs.~\cite{Kusenko:2006rh,Petraki:2007gq}, we consider a real SU(2) singlet scalar $S$ which couples to the Higgs sector and acquires a vacuum expectation value (VEV) after spontaneous symmetry breaking (SSB). The SM Lagrangian extended to include the sterile neutrino fields $N_a$ and the singlet Higgs $S$, becomes
\be
\mathcal{L} = \mathcal{L}_{\rm{SM}} + i \bar{N}_a\not{\partial}N_a - y_{a \alpha} H^{\dagger}\bar{L}_{\alpha} N_a - \frac{f_a}{2} S \bar{N}^{c}_{a}N_{a} - V(H,S) + h.c.,
\label{lagr}
\ee
where the most general gauge-invariant, renormalizable scalar potential is
\be
V(H,S) = -\mu_{H}^{2} |H|^{2} -\frac{1}{2} \mu_{S}^{2}S^{2} + \frac{1}{6} \alpha S^{3} + \omega |H|^{2} S + \lambda_{H} |H|^{4} + \frac{1}{4} \lambda_{S} S^{4} + 2 \lambda_{HS} |H|^{2} S^{2}.
\label{V}
\ee
Upon the electroweak symmetry breaking (EWSB), both the singlet and the neutral component of the doublet Higgs acquire VEVs.
This gives rise to the Majorana masses of sterile neutrinos
\be
\MN = \fN \langle S \rangle,
\label{Majorana}
\ee
but it also leads to mass mixing between the two scalar fields. As a result, both Higgs mass eigenstates will couple to the SM particles and the sterile neutrinos.

The collider implications of an effective coupling of sterile neutrinos to the SM Higgs have been explored by Graesser~\cite{Graesser:2007yj,Graesser:2007pc}, who considered the lowest order non-renormalizable interaction
\be
\delta \mathcal{L} = \frac{c}{\Lambda} NN H^{\dagger} H.
\label{hhnn}
\ee
After EWSB, the vertex of Eq.~(\ref{hhnn}) allows the SM Higgs to decay to sterile neutrinos provided that such a decay is kinematically open. The model we present here can be considered as a particular UV completion of the class of effective theories encompassed by Eq.~(\ref{hhnn}). However, the introduction of explicit renormalizable interactions which generate \eq{hhnn} results in new effects and distinct phenomenology.
%Such effects become important, since the scale of new physics described here will be probed by the upcoming collider experiments.
In our discussion below, we incorporate, in the appropriate parametric regime, many of the signatures arising from the Higgs -- sterile neutrino coupling,
pointed out in Refs.~\cite{Graesser:2007yj,Graesser:2007pc}, and extend the analysis to include the new observable signals of our model.
Signatures of sterile-neutrino coupling to the Higgs sector have also been considered in the context of the Majoron models~\cite{Aranda:2007dq,Cline:2009sn}.

The mixing of the doublet Higgs with a singlet scalar suppresses the gauge interactions of the Higgs field, and relaxes the experimental limits on the Higgs boson mass~\cite{O'Connell:2006wi,BahatTreidel:2006kx,Barger:2006sk,Barger:2007im}, thus reducing the tension between the LEP lower bound and the upper bound arising from electroweak precision observables (EWPO). The effect is enhanced by the coupling of the Higgs sector to gauge-singlet fermions, due to the new decay channel introduced.
The existence of heavy sterile neutrinos would imply that the Higgs boson can be significantly lighter than what anticipated in the SM.

The number of sterile neutrinos is not constrained by any theoretical arguments, such as anomaly cancellation. Although, at least two sterile neutrinos are needed in order to produce the observed active neutrino mass splittings. The contribution of singlet neutrinos to the active neutrino masses comes in a combination of their Majorana mass $M_{_{\rm{N}}}$, and their mixing
$\theta^2 \sim y^2 \langle H \rangle^2 /\MN^2$ to the left-handed species. Sterile neutrinos for which $M_{_{\rm{N}}} \sin^2 \theta \approx 2\times 10^{-2} \eV$, have a dominant contribution to the active neutrino mass matrix. Smaller contributions can produce subleading structures in the mass matrix, while larger induced masses cannot be excluded since detailed cancelations may occur~\cite{Smirnov:2006bu}.

Here, for simplicity we consider only one sterile-neutrino species coupled to the singlet Higgs. Our focus will be on sterile neutrinos which can contribute significantly to the active neutrino masses.  Since the see-saw mechanism can accommodate the observed neutrino masses for a variety of Majorana scales and active-sterile mixings, we will merely focus on scales which can be probed experimentally.  Existing limits on sterile-neutrino masses and mixings push the region of interest at masses $\MN \gtrsim 1 \GeV$ and mixings $\sin^2 \theta \lesssim 10^{-10}$~\cite{Smirnov:2006bu}. Interestingly, such masses can arise via SSB at the EW scale for reasonably large sterile-neutrino Yukawa couplings to the singlet Higgs, $\fN \gtrsim 10^{-3}$. This makes these particles potentially discoverable at LHC, despite their rather tiny mixing with active neutrinos.

Moreover, if a keV sterile neutrino acquires its mass via an EW-scale singlet Higgs it can be produced via decays of the singlet scalar with the correct abundance to account for the dark matter of the universe~\cite{Kusenko:2006rh,Petraki:2007gq}. The presence of a singlet scalar in the Higgs sector opens the possibility for a 1st order EW phase transition~\cite{O'Connell:2006wi,Barger:2007im,Profumo:2007wc,Petraki:2007gq}, which enhances the prospects for EW baryogenesis. The simple extension of the SM described here can thus accommodate the observed relic matter density, the neutrino masses and the matter-antimatter asymmetry, while also providing observable signatures for collider experiments.

The scenario most distinct from SM phenomenology in our model will be when the doublet-like Higgs has substantial branching fraction into $NN$.  Thus the signals arising from the decay of sterile neutrinos will be important in assessing the discoverability of the model.  Signals arising from sterile neutrinos that have been studied in a number of schemes~\cite{Gorbunov:2007ak,Graesser:2007yj,Graesser:2007pc,Aranda:2007dq,Perez:2009mu,Han:2006ip,Bezrukov:2009th}. We summarize the relevant signatures in section~\ref{Sec N Decays}.

This paper is organized as follows: In section~\ref{Sec H Sector} we discuss the extended Higgs sector of \eq{V}. We explore the effects of Higgs mixing and the signatures that arise from decays of the Higgs bosons. We present our results with specific illustrative model examples. We show how the experimental lower limit on the Higgs mass is modified due to the doublet-singlet mixing in the Higgs sector, and how the new bounds vary with the sterile-neutrino coupling to the singlet Higgs.
In section~\ref{Sec N Decays} we present the signatures that arise from sterile-neutrino decays produced in colliders due to their coupling to the Higgs sector.
We conclude in section~\ref{Sec concl}.

%%%%
\section{The extended Higgs Sector: Signatures and Bounds \label{Sec H Sector}}
%%%%

\subsection{The scalar potential \label{potential}}

After SSB, the neutral component of the doublet Higgs $H^{0} = (v + h)/\sqrt{2}$, and the singlet Higgs $S = \sigma + s$,  acquire VEVs and mix in the scalar potential, to produce two mass eigenstates $H_1, \: H_2$
\bea
\left(
\begin{array}{c}
H_1 \\
H_2
 \end{array}
\right) =
\left(
\begin{array}{lr}
~~ \cos \phi & \sin \phi \\
 - \sin \phi & \cos \phi
 \end{array}
 \right)
 \left(
 \begin{array}{c}
 h \\
 s
 \end{array}
 \right).
\eea

The scalar potential of Eq.~(\ref{V}) contains 7 parameters, 4 of which are dimensionful and 3 dimensionless. At tree level, they correspond equivalently to the two Higgs mass eigenvalues $M_1, \: M_2$, the two VEVs $v, \: \sigma$, the mixing angle $\phi$, and two dimensionless constants $\kappa, \: \lambda$ which determine the behavior of the potential away from the minimum. Imposing the standard doublet VEV, $v = 246 \GeV$, leaves 6 free parameters.

In what follows, we take $H_1$ to be always lighter than $H_2$, i.e. $M_1 < M_2$. The mixing $\phi$ determines the coupling of the Higgs states to the SM particles and the sterile neutrinos. Whether the lightest Higgs state is more doublet-like, or singlet-like is, of course, critical for the signatures obtained from Higgs decays, as we will discuss in Sec.~\ref{Sec H Decays}.
Moreover, we assign $(2 \kappa M_2)$ and $\lambda$ as the $H_1^2 H_2$ and $H_1^3 H_2$ couplings respectively. These terms arise after SSB and allow decay of the heavier mass eigenstate $H_2$ into two or three light Higgs particles $H_1$, provided that such decays are kinematically allowed. Such decay channels can provide powerful signatures of the extended Higgs sector considered here. We discuss them in Sec.~\ref{Sec H Decays}.

Although it is straightforward to obtain the scalar potential in terms of $H_1, \: H_2$ and the new parameters introduced, we will not present it explicitly here, as many of the terms that arise after SSB appear somewhat messy. It suffices, though, to state the terms which will be important in our discussion
\be
V(H_1,H_2) \: \supset \: \frac{1}{2} M_1^2 H_1^2 + \frac{1}{2} M_2^2 H_2^2 + 2\kappa M_2 \: H_1^2 H_2 + \lambda H_1^3 H_2
\label{Vnew}
\ee
Cubic and quartic couplings that arise after SSB and are not included in \eq{Vnew} are functions of the independent parameters $M_1, M_2, v, \sigma, \phi, \kappa, \lambda$.
These terms may affect the expected Higgs signal significances at LHC, as studied in~\cite{BahatTreidel:2006kx,Barger:2006sk}, but they yield subdominant corrections to the Higgs decay rates, which we will now discuss.

\subsection{Decays of the Higgs bosons \label{Sec H Decays}}

Due to the mixing in the scalar potential, both Higgs bosons will effectively couple and decay to sterile neutrinos and SM particles. The scalar interactions between the two Higgs states may also induce decay of one into another. In this section we describe the decays of the Higgs bosons. Since the non-SM couplings relax existing limits on the Higgs mass, we consider a wider range of Higgs masses than what allowed in the SM. We postpone the discussion on how the Higgs mass bounds are modified until Sec.~\ref{Sec LEP}. The Higgs decays we describe here have to be considered in conjunction with the sterile-neutrino decay modes and signatures, which we analyze in Sec.~\ref{Sec N Decays}.

The coupling of the Higgs bosons $H_1, H_2$ to the sterile neutrino $N$ is determined by the sterile-neutrino Yukawa coupling $\fN$ and the Higgs mixing $\phi$
\be
\foneN = \fN \sin \phi,  \quad \ftwoN =  \fN \cos \phi.
\label{fs}
\ee
The partial widths for Higgs decay into sterile neutrinos, $H_i \rightarrow N N$, are
\be
\Gamma_{i,N} = \fiN^2 \: \frac{M_i}{8 \pi} \left( 1 - \frac{4 \MN^{2}}{M_i^2}\right)^{3/2} .
\label{Gamma H-NN}
\ee

The cubic and quartic couplings in the scalar potential of \eq{V} allow for the possibility of the heavy Higgs state $H_2$ to decay into two or three light Higgs particles $H_1$. If such decays are kinematically accessible, the corresponding rates are
\be
\Gamma(H_2 \rightarrow 2 H_1 ) = \frac{\kappa^2 M_2}{8 \pi} \left( 1 - \frac{4 M_1^2}{M_2^2} \right)^{1/2},
\label{2H1}
\ee
and to a good approximation
\be
\Gamma(H_2 \rightarrow 3 H_1 ) \simeq \frac{3\lambda^2 M_2}{256 \pi^3} \left( 1 - \frac{9 M_1^2}{M_2^2} \right)^{5/2}.
\label{3H1}
\ee

The coupling of the Higgs doublet to the various SM particles, $g^{\rm SM}_j h \: X_j X_j$,
yields the couplings of the two Higgs mass eigenstates to the SM fields
\be
g_{1,j} \equiv g^{\rm SM}_j \cos \phi, \quad g_{2,j} \equiv g^{\rm SM}_j \sin \phi \label{gs}
\ee
The partial decay widths of $H_1, H_2$ into SM states can be found from those of the SM Higgs, by substituting $g^{\rm SM}_j \rightarrow g_{i,j}$.

\bigskip

The branching ratios of $H_1$ and $H_2$ to SM particles are altered in respect to the SM predictions, due to:
\begin{enumerate}
\renewcommand{\theenumi}{\roman{enumi}}
 \item The mixing suppression on the Higgs -- SM couplings, according to \eq{gs}.
 \item Any new decay modes that become available. Both Higgs bosons may decay into sterile neutrinos, as described by Eqs.~(\ref{fs}) and (\ref{Gamma H-NN}). The heavy Higgs state $H_2$ may also decay into light Higgs particles $H_1$, if kinematically allowed, according to Eqs.~(\ref{2H1}) and (\ref{3H1}).
\end{enumerate}
Thus, the new branching ratios for the $H_i \rightarrow X_j \: X_j$ decays become suppressed according to
\be
{\rm Br} (H_i \rightarrow X_j \: X_j) = \left.{\rm Br}^{\rm SM}_j\right/ (1 + A_i)  %\frac{{\rm BR}^{\rm SM}_j}{1 + A_i}
\label{BR}
\ee
where ${\rm Br}^{\rm SM}_j$ is the SM branching ratio for decay of the Higgs into $X_j \: X_j$, and
\be
A_1 \equiv \frac{\Gamma_{1,N}}{\cos^2 \phi \:\: \Gamma^{\rm SM}_{\rm tot}} \quad  {\rm and} \quad
A_2 \equiv \frac{\Gamma_{2,N} + \Gamma_{H_2 \rightarrow 2 H_1} + \Gamma_{H_2 \rightarrow 3 H_1} }{\sin^2 \phi \:\: \Gamma^{\rm SM}_{\rm tot}}
\label{A}
\ee
with $\Gamma^{\rm SM}_{\rm tot}$ being the SM Higgs total decay rate.
In the limit that $A_i \ll 1$, the decay branching fractions into SM particles reduce to that of the SM Higgs. In the opposite limit the branching ratios to SM modes are strongly suppressed, and the non-SM decay channels dominate.
The decay branching ratios via all non-SM channels are
\be
{\rm Br} (H_i \rightarrow {\rm non-SM}) = A_i/(1+A_i)
\label{BR-non SM}
\ee

We note here, that Eqs.~(\ref{BR}) and (\ref{A}) for the lightest Higgs state $H_1$, parallel the predictions of Refs.~\cite{Graesser:2007pc,Graesser:2007yj}, where the non-renormalizable interaction of \eq{hhnn} was considered, if we formally correspond $\fN^2 \tan^2 \phi \rightarrow (c v / \Lambda)^2$. However, the introduction of a second Higgs state, and the possibility for the heaviest of the Higgs bosons to decay into the lighter one, render the signatures of our model quite distinct. Moreover the present model has distinct decay rates and one extra degree of freedom: the real singlet Higgs.

In the SM, if the Higgs mass is below the $W^+W^-$ production threshold, $\sim 160 \GeV$, the total decay width of the Higgs boson is dominated by the $\bar{b}b$ channel. In this regime, $\Gamma^{\rm SM}_{\rm tot} \simeq 3 \left(g^{\rm SM}_b \right)^2 M_h /8 \pi$, and:
\bea
A_1 &\simeq&  \frac{1}{3} \left(\frac{\fN}{g^{\rm SM}_b}\right)^2 \tan^2 \phi \label{A1-small M} \\
A_2 &\simeq&  \frac{\fN^2 \cot^2 \phi + (\kappa^2 + 3\lambda^2/32 \pi^2) \csc^2 \phi}{3 \left(g^{\rm SM}_b\right)^2}  \label{A2-small M}
\eea
where $g^{\rm SM}_b \simeq 2 \cdot 10^{-2}$.
Evidently, in order for a doublet-like or even a doublet-singlet maximally-mixed Higgs state in our model to yield a strong sterile-neutrino signature (assuming this is kinematically allowed), the sterile-neutrino Yukawa coupling has to be rather large, $\fN~\gtrsim~10^{-2}$.

If the Higgs mass is above the $W^+W^-$ production threshold, the decay into weak gauge bosons prevails over the fermionic decay modes, and $\Gamma^{\rm SM}_{\rm tot} = 3 M_h^3/ (32 \pi v^2)$. In this case, the suppression factors $(1+A_i)^{-1}$ for decay into SM particles depend on the Higgs boson masses:
\bea
A_1 &\simeq&  \frac{4}{3}  \frac{v^2}{M_1^2} \: \fN^2 \tan^2 \phi \label{A1-large M} \\
A_2 &\simeq&  \frac{4}{3}  \frac{v^2}{M_2^2} \: \left[ \fN^2 \cot^2 \phi + (\kappa^2 + 3\lambda^2/32 \pi^2) \csc^2 \phi  \right] \label{A2-large M}
\eea
Eqs.~(\ref{A1-large M}) and (\ref{A2-large M}) imply that if the Higgs mass is sufficiently large, its decay will exhibit the familiar SM branching ratios. However, for a considerable range of Higgs masses, it is also possible that the Higgs bosons decay at comparable rates via both SM and non-SM modes.

We note here that the Higgs couplings to sterile neutrinos and SM particles, Eqs.~(\ref{fs}) and (\ref{gs}), satisfy the signature relations
\bea
\foneN^2 + \ftwoN^2   &=& \fN^2,                        \label{fs square} \\
g_{1,j}^2 + g_{2,j}^2 &=& \left(g^{\rm SM}_j \right)^2, \label{gs square}
\eea
which translate into a relation between $A_1$ and $A_2$, or equivalently between the branching ratios of the two Higgs states.
It is also obvious from Eqs.~(\ref{A1-small M}) -- (\ref{A2-large M}), that the exact value of the singlet VEV $\sigma$ does not affect the Higgs decay branching ratios, except it determines the lowest Higgs mass at which decay into sterile neutrinos becomes kinematically accessible.

\bigskip

\begin{table}[t]
\begin{center}
\begin{tabular}{| l |  c | c | c | }
\hline
              & Model A      & Model B & Model C \\ \hline \hline
$\sin \phi$   & $1/\sqrt{2}$ & 0.01    & 0.99    \\ \hline
Higgs states  & $H_{1,2} = \frac{1}{\sqrt{2}}\left(h \pm s \right)$
              & $H_1 \approx h$, $H_2 \approx s$
              & $H_1 \approx s$, $H_2 \approx h$ \\ \hline
LEP constraints   &  $M_{1,2} \geqslant 80 \GeV$   &  $M_{1,2} \geqslant 114 \GeV$    & $ M_2 \geqslant 114 \GeV$ \\ \hline
EWPO constraints  &  $M_{1,2} \leqslant 220\GeV$   &  $M_1 \leqslant 185 \GeV$    & $ M_{1,2} \leqslant 185 \GeV$ \\ \hline
\end{tabular}
\end{center}
\caption{Representative scenarios of singlet-doublet Higgs mixing, and experimental bounds on the masses of the Higgs bosons.
In Model A, the mixing is maximal, and the couplings of the light and heavy Higgs bosons to particles are equal.
(The LEP bounds assume $\fN = 10^{-1}$).
In Model B, the light (heavy) Higgs is dominantly doublet (singlet). The light Higgs behaves effectively as the SM Higgs.
In Model C, the light (heavy) Higgs is dominantly singlet (doublet).}
\label{models}
\end{table}

\begin{figure}[t]
\centering
\begin{tabular}{cc}
\epsfig{file=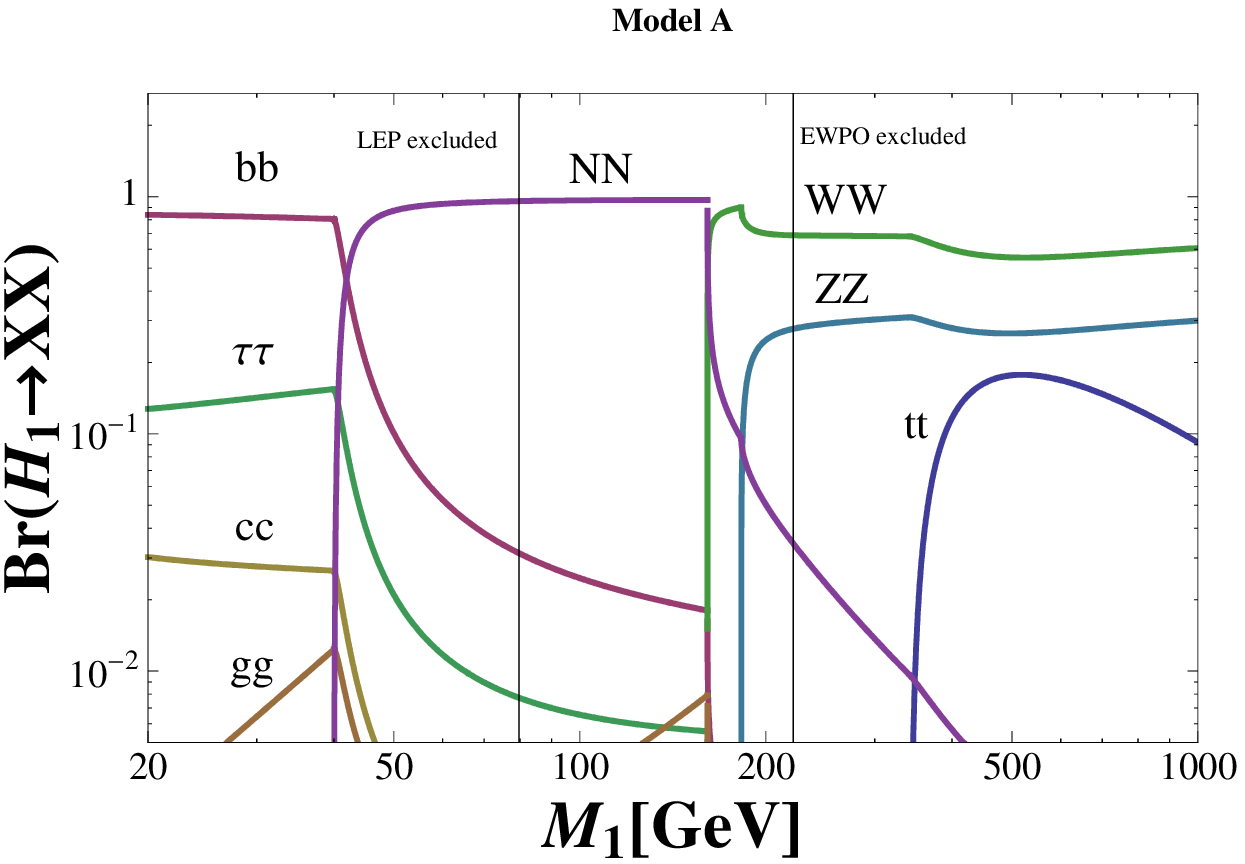,width=0.5\linewidth,clip=}
\epsfig{file=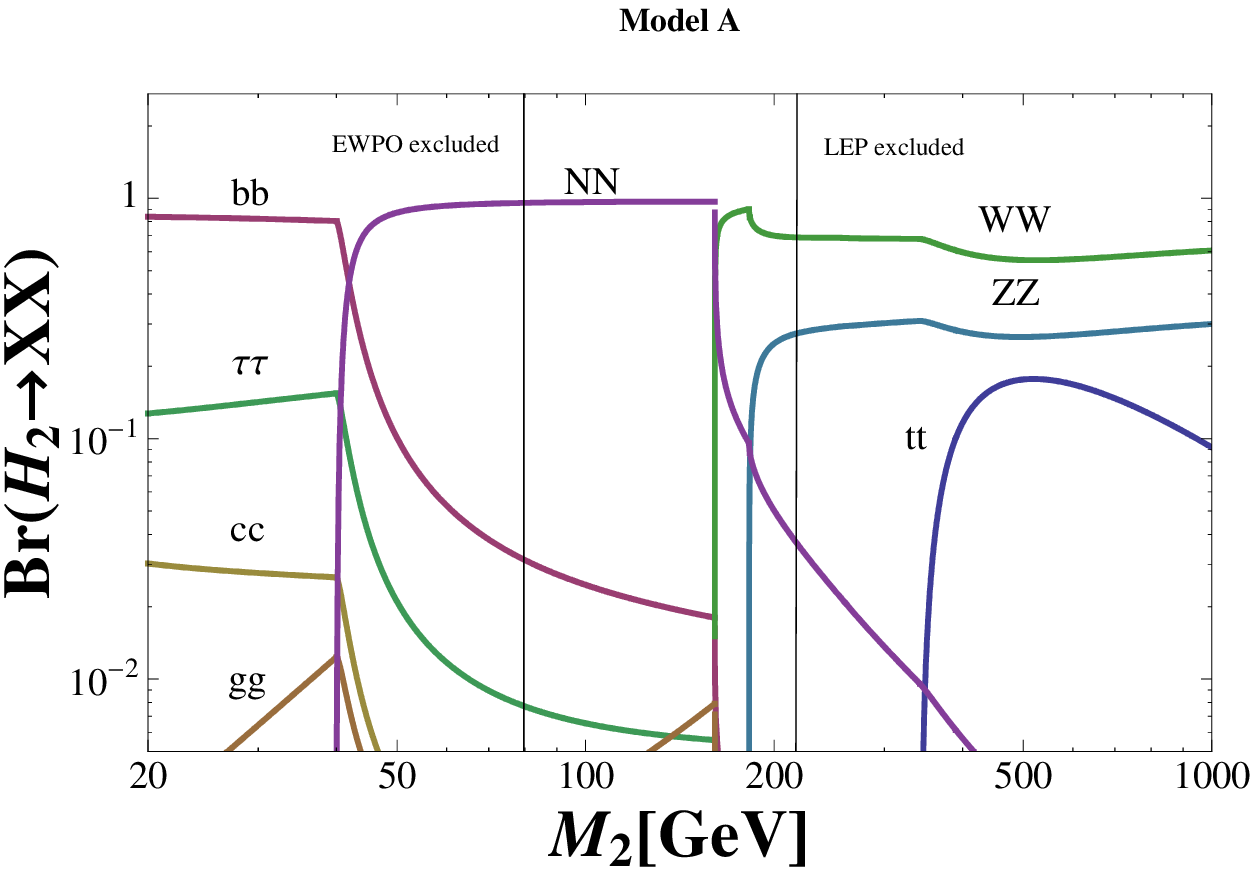,width=0.5\linewidth,clip=}\\
\\
\epsfig{file=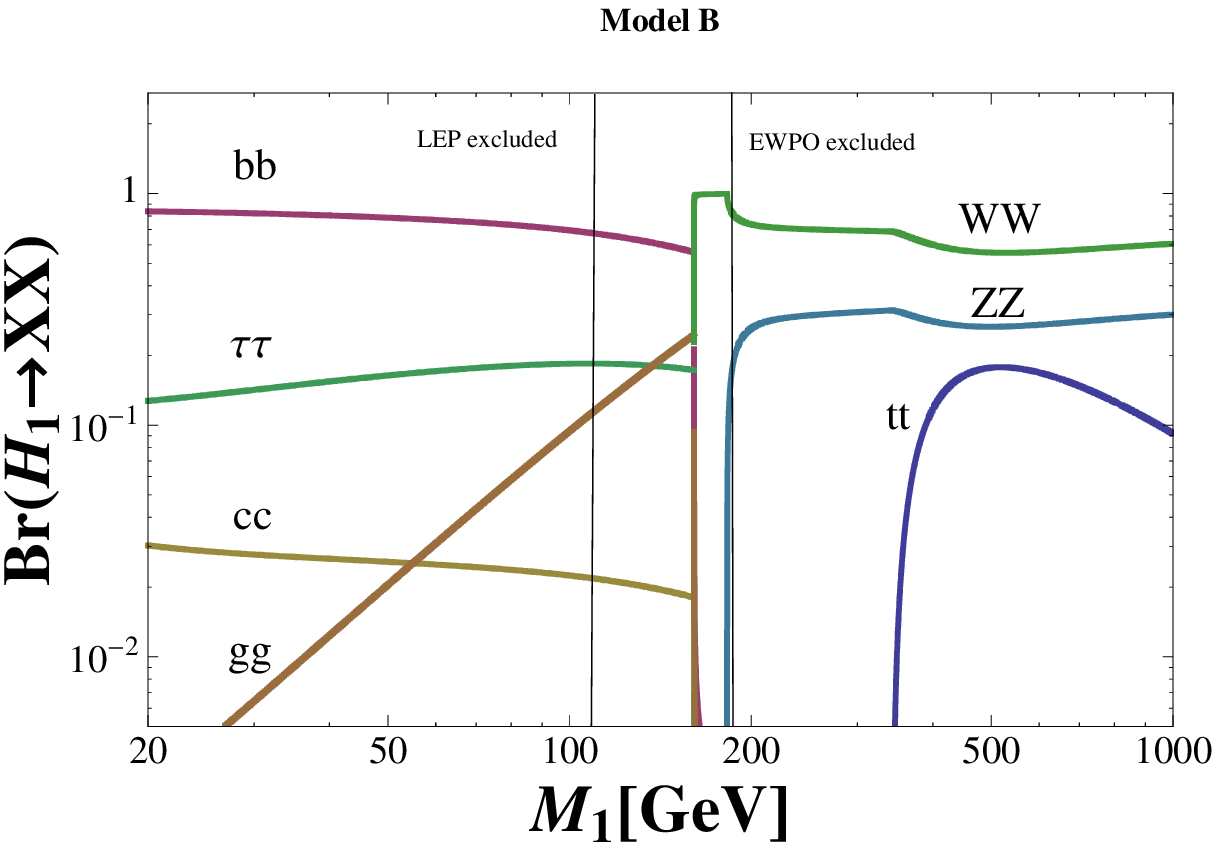,width=0.5\linewidth,clip=}
\epsfig{file=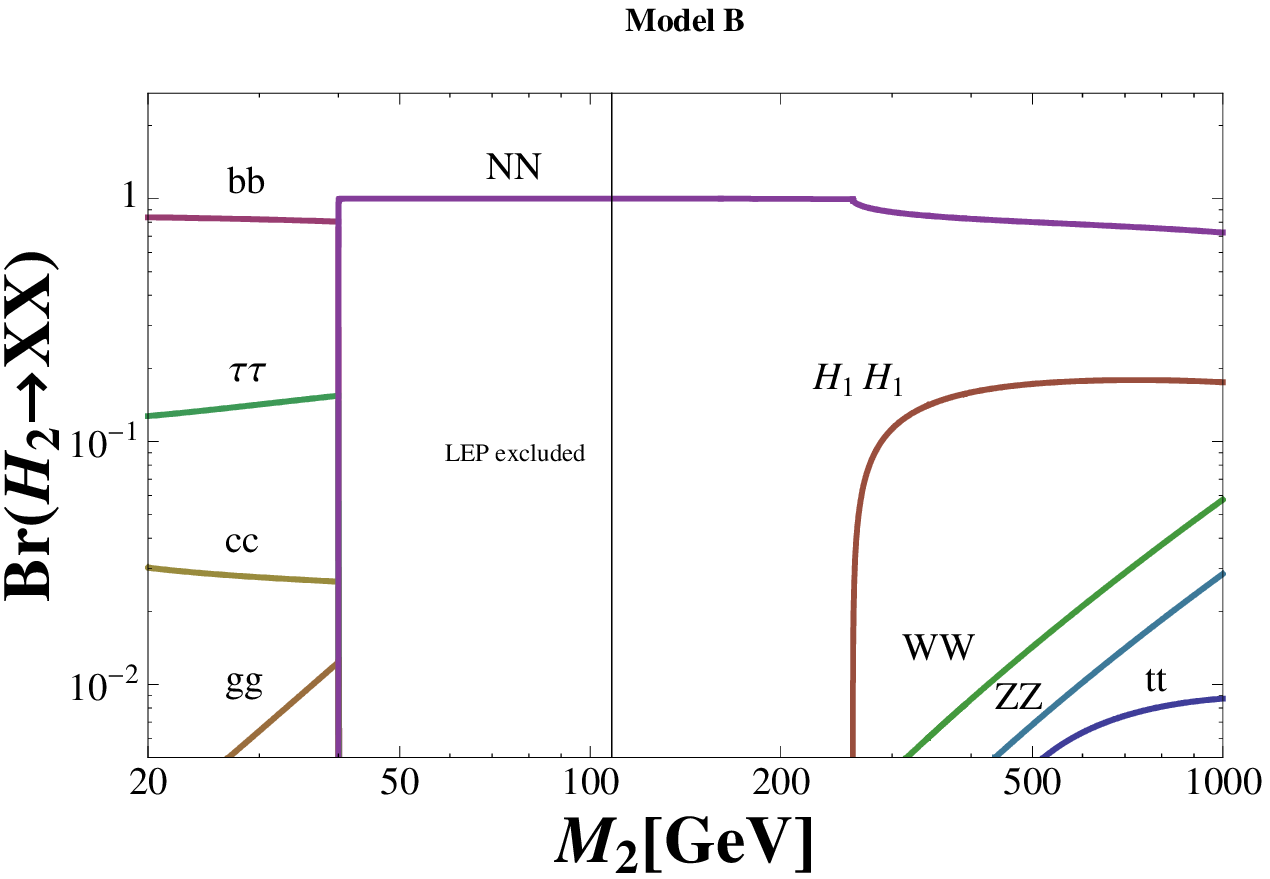,width=0.5\linewidth,clip=} \\
\\
\epsfig{file=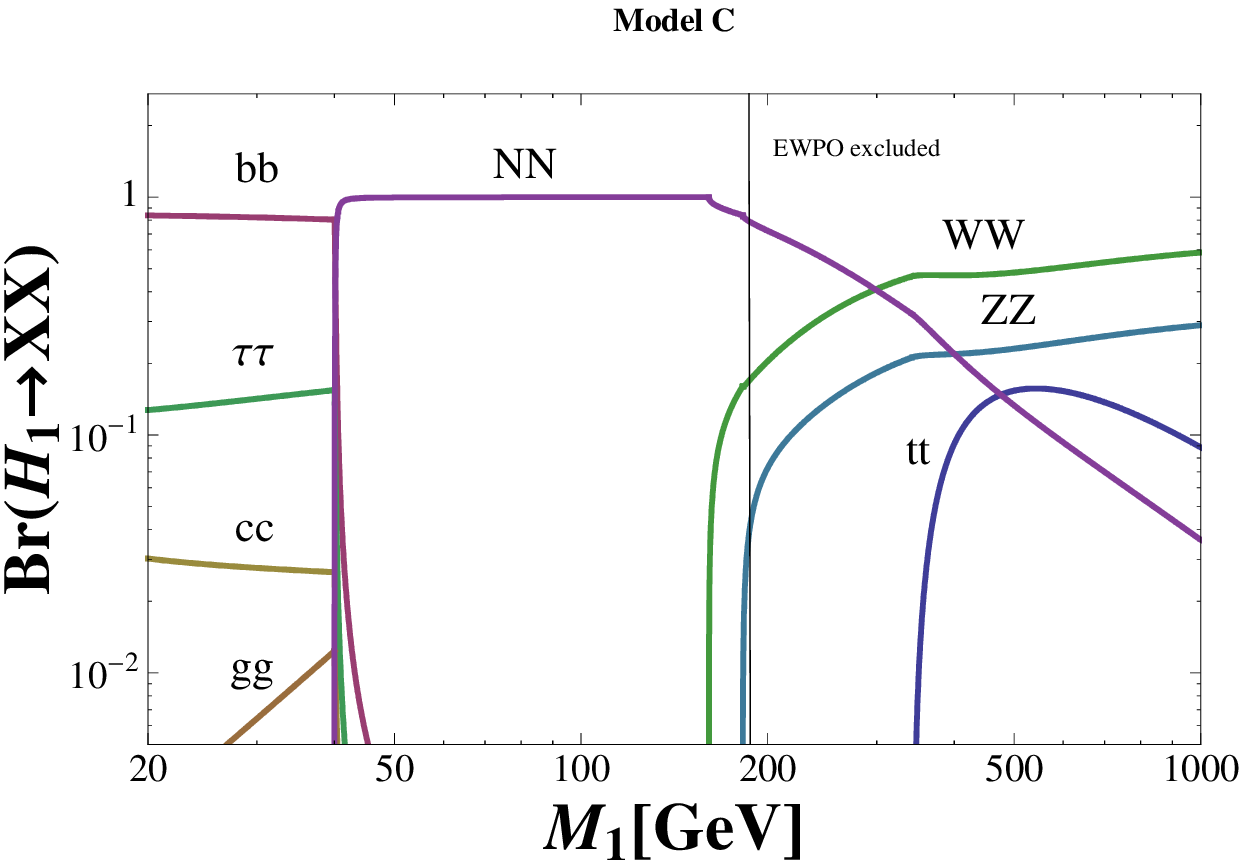,width=0.5\linewidth,clip=}
\epsfig{file=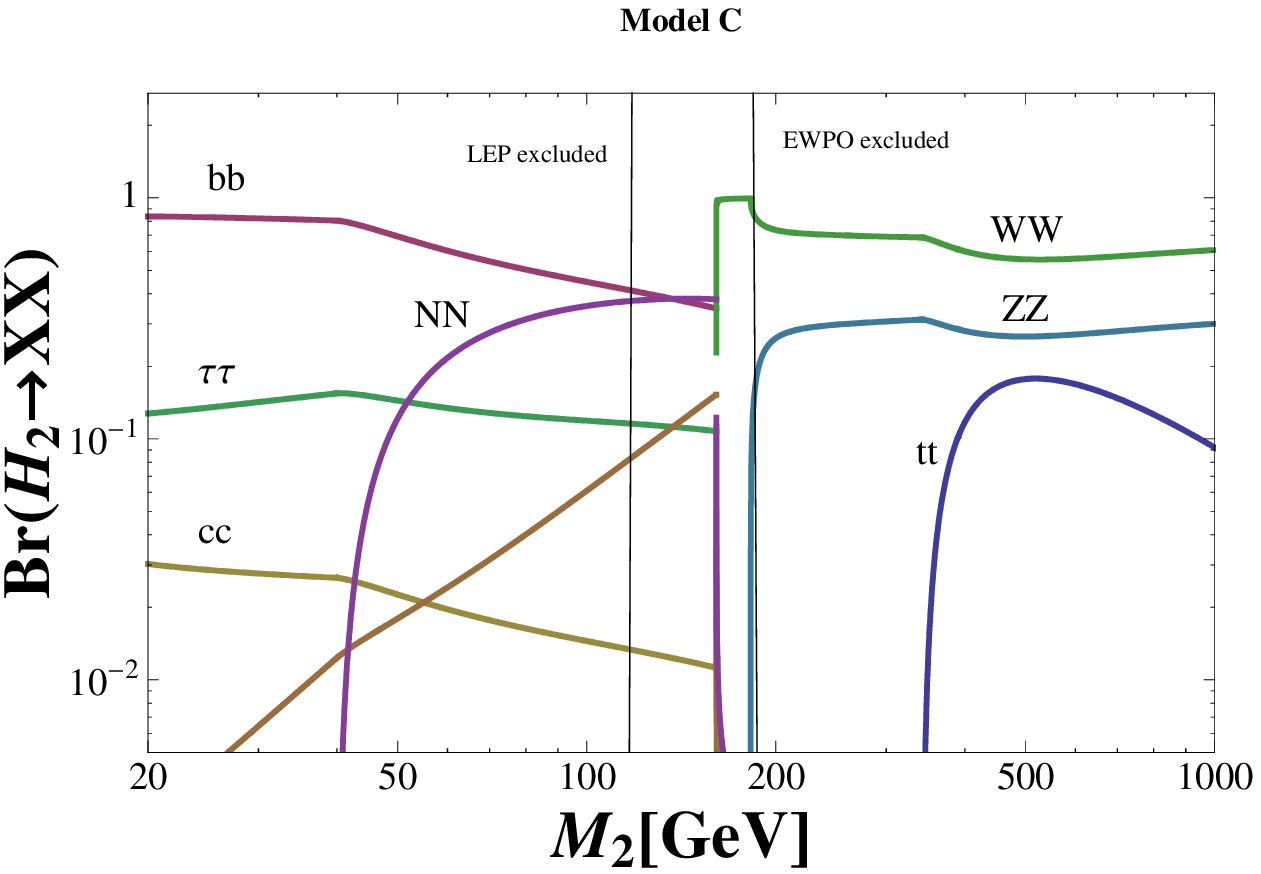,width=0.5\linewidth,clip=} \\
\end{tabular}
\caption{The decay branching ratios of the two Higgs mass eigenstates, $H_1$ (left column) and $H_2$ (right column), vs their mass, for various values of the doublet-singlet Higgs mixing. In model A $\sin \phi = 1/\sqrt{2}$; in model B $\sin \phi = 0.01$; in model C $\sin \phi = 0.99$.
We have assumed $\fN = 0.1$ and $\sigma = 200 \GeV$, resulting in a sterile neutrino of mass $\MN = 20 \GeV$. For the $H_2$ plots, we have taken $M_1 = 130~\rm{GeV}$ which is consistent with LEP and EWPO in all cases, $\kappa = 0.05$, and $\lambda = 0.1$.
Also shown, the LEP lower bounds and the EWPO upper bounds on the mass of the lightest Higgs, for each of the models considered. The regions on the left of the LEP lines, and the regions on the right of the EWPO lines are excluded.
}
\label{H graphs}
\end{figure}

We will now sketch the above by considering three illustrative sample models. We fix the sterile-neutrino Yukawa coupling to the Higgs sector at $\fN = 10^{-1}$, and in models A, B and C, we examine the Higgs decay channels for three different values of the Higgs mixing, as shown in Table~\ref{models}. The branching ratios of the Higgs states $H_1$ and $H_2$, for a wide range of masses, are presented in Fig.~\ref{H graphs}.

Model $A$ corresponds to maximal mixing between the singlet and the doublet Higgs bosons, $\tan \phi =1$. The decays of the two Higgs mass eigenstates will thus be similar. If the sterile-neutrino Yukawa coupling is large $\fN \gtrsim 10^{-2}$, as assumed in the plots of Fig.~\ref{H graphs}, the dominant decay of the Higgs bosons will be into sterile neutrinos, provided that this channel is kinematically accessible, and as long as the Higgs masses do not exceed the threshold for $W^+ W^-$ production. For Higgs masses $M_{1,2} \gtrsim 160 \GeV$, the decay into weak gauge bosons dominates. Nonetheless, it is possible that the heavy Higgs has a non-negligible decay branching ratio into light Higgs bosons, ${\rm Br} (H_2 \rightarrow 2H_1) \gtrsim 10^{-3}$, if $\kappa \gtrsim 10^{-2}$.

In Model B, the lightest Higgs boson is mostly doublet. The branching ratios of $H_1$ follow the SM predictions, while $H_2$ will decay dominantly into sterile neutrinos, if this is kinematically possible. Due to its singlet-like character, the decay of $H_2$ into sterile neutrinos may dominate even over the $WW$ mode, for a wide range of masses. For the same reasons, the decay of $H_2$ into two or three $H_1$ particles can occur at a significant fraction.

In Model C, the light Higgs is mostly singlet, and can thus provide a strong sterile-neutrino signal. The branching ratios into sterile neutrinos and gauge bosons can be comparable for a significant range of Higgs masses. The heaviest Higgs exhibits SM-like behavior.

\subsection{Modified bounds on the Higgs boson mass \label{Sec LEP}}

The LEP lower bound on the mass of the Higgs boson, $M_h > 114~\GeV$~\cite{Barate:2003sz}, is deduced from the $e^+e^- \rightarrow HZ$ searches and the Higgs decay into SM final states.
However, smaller Higgs masses are still allowed provided that the Higgs production channel and/or its decay into SM particles are sufficiently suppressed.
The LEP data can then be employed to obtain limits on non-standard couplings of the Higgs sector to the SM particles.
In particular, it can constrain the ratio of the non-SM over SM Higgs-production cross-sections, weighted by the Higgs decay branching fraction into SM particles (see e.g. Ref.~\cite{Barger:2007im}). The relevant quantity $\xi^2$ for the two Higgs states of the model considered here is
\be
\xi_i^2 = \left\{
\begin{array}{lr}
\left(\frac{g_{H_1 ZZ}}{g_{HZZ}^{\rm SM}}\right)^2 \frac{ \Gamma^{\rm SM}_{\rm tot}}{\Gamma^{\rm SM}_{\rm tot}  + \Gamma (H_1 \rightarrow N N)}, &  i = 1\\
\left(\frac{g_{H_2 ZZ}}{g_{_{HZZ}}^{\rm SM}}\right)^2 \frac{ \Gamma^{\rm SM}_{\rm tot}}{\Gamma^{\rm SM}_{\rm tot}  + \Gamma (H_2 \rightarrow N N) + \Gamma(H_2 \rightarrow 2H_1) + \Gamma(H_2 \rightarrow 3H_1)}, &  i = 2
\end{array}
\right.
\label{xi}
\ee
or, in the notation introduced in the previous section
\be
\xi_1^2 = (1+A_1)^{-1} \cos^2 \phi \quad  {\rm and} \quad \xi_2^2 = (1+A_2)^{-1} \sin^2 \phi
\label{xi_simpl}
\ee

\begin{figure}
\centering
\epsfig{file=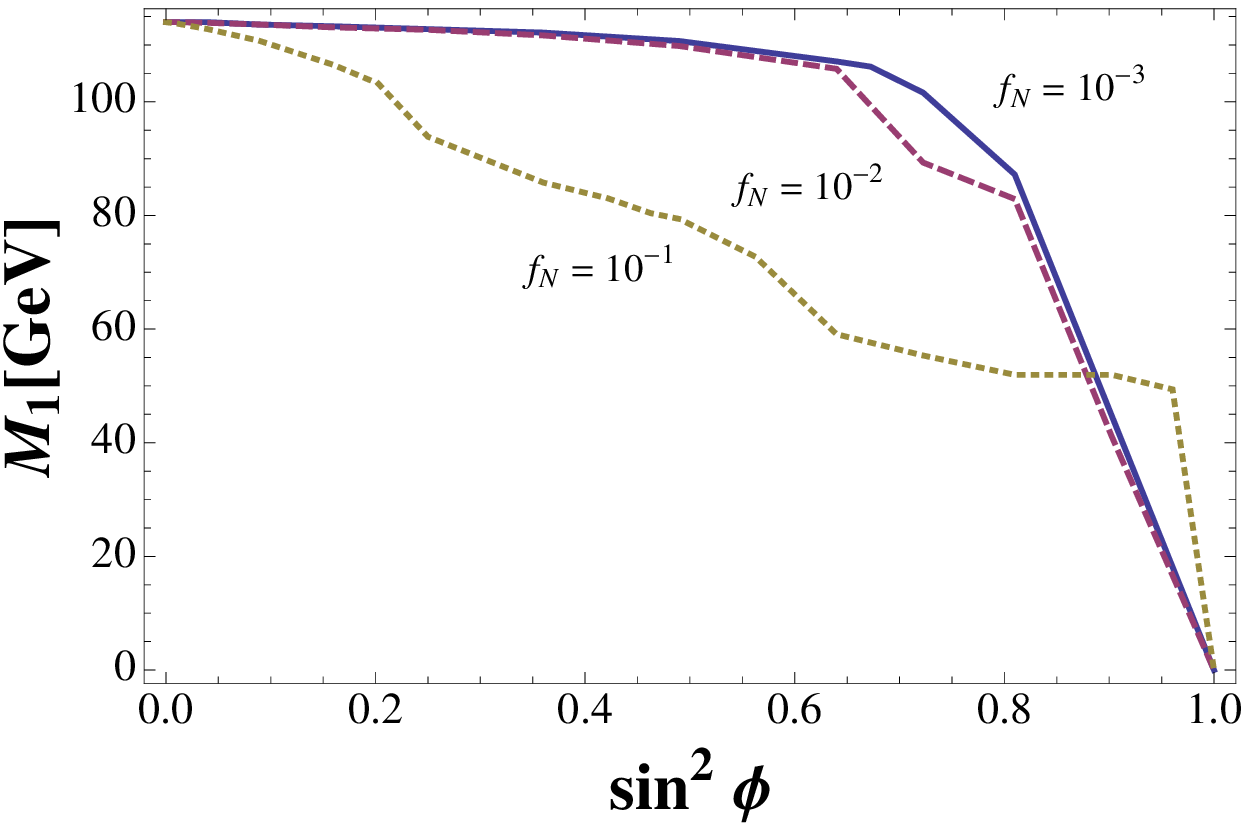,width=0.6\linewidth,clip=}
\caption{The lower bound on the mass $M_1$ of the lightest Higgs state $H_1$ based on LEP data~\cite{Barate:2003sz}, vs the doublet-singlet Higgs mixing angle $\phi$, for various values of the sterile-neutrino Yukawa coupling to the singlet Higgs. The curves shown correspond to $\fN = 10^{-3}$ (solid, upper, blue line), $\fN = 10^{-2}$ (dashed, middle, purple line), $\fN = 10^{-1}$ (dotted, lower, yellow line). We have assumed $\sigma = 200 \GeV$, which gives rise to sterile neutrinos of mass 0.2~GeV, 2~GeV, and 20~GeV respectively.
The regions below the curves are excluded.
}
\label{LEP}
\end{figure}

We summarize the LEP limits on the light Higgs $H_1$ in Fig.~\ref{LEP}.
We have used the maximum of the observed bound and the background expectation at 95\% C.L.~\cite{Barate:2003sz}, as the upper limit on $\xi^2$.
The resulting mass bound set by the LEP data depends both on the Higgs mixing, and on the sterile-neutrino coupling to the Higgs sector. In the limit $\sin \phi \rightarrow 0$, the singlet sector (scalar + fermions) decouples, and the SM lower mass bound on the Higgs mass, 114~GeV, is recovered. In the opposite limit, $\sin \phi \rightarrow 1$, the lightest Higgs is mostly singlet, and the lower bound on the Higgs mass disappears, since essentially none of the light Higgs particles are produced at LEP.

The limit on the Higgs mass is relaxed as the sterile-neutrino coupling to the Higgs sector increases, since the production of Higgs bosons at LEP may be concealed due to their decay into the singlet fermions. This is true as long as the $H_1 \rightarrow NN$ decay is kinematically open, i.e. for $M_1 \geqslant 2 \fN \sigma$. For lower masses $M_1$, the suppression of the Higgs-mass lower bound is only due to the mixing in the Higgs sector. This gives rise to the line crossing observed in Fig.~\ref{LEP}. Additional sterile neutrinos would further relax the Higgs lower mass bound.

Although we do not present the modified LEP bounds for the heavier Higgs state $H_2$, we note that the lower limit on $M_2$ can be obtained from Fig.~\ref{LEP}, if we correspond $\sin^2 \phi \rightarrow \cos^2 \phi$. This holds as long as the decays $H_2 \rightarrow 2 H_1 \: {\rm or} \: 3 H_1$, Eqs.~(\ref{2H1}) and (\ref{3H1}), are negligible. If such decay modes become important, the limit on $M_2$ is further relaxed.

\bigskip

In the SM, the Higgs mass is constrained by EWPO to be $M_h \lesssim 185 \GeV$. If a singlet scalar couples to the Higgs sector, it will appear in the gauge-boson propagators at one-loop level, and can thus affect the EWPO. This was considered in Ref.~\cite{Barger:2007im}, where upper bounds for the masses of the two Higgs eigenstates were derived.
If the singlet Higgs also couples to sterile neutrinos, additional contributions arise only at the two-loop level. This effect is negligible, and the EWPO constraints in our model are very nearly the same as the bounds derived in Ref.~\cite{Barger:2007im}.

\bigskip

In the case of maximal mixing, as in Model A, the EWPO constraints imply $M_{1,2} \lesssim 220 \GeV$~\cite{Profumo:2007wc,Barger:2007im}. The LEP bound is relaxed to $M_{1,2} \gtrsim 80 \GeV$, for $\fN = 10^{-1}$, as seen from Fig.~\ref{LEP}.
In Model B, the lightest Higgs is dominantly doublet, and behaves essentially as a SM Higgs, with vanishingly small branching fraction to sterile neutrinos. It is thus constrained by both the LEP and the EWPO bounds for the SM Higgs.
In Model C, the doublet-like, heavy Higgs behaves as in the SM, and is similarly constrained. The LEP bound for the lightest, singlet-like Higgs vanishes.
The LEP and EWPO bounds on the masses of the Higgs bosons, for the three models considered, are summarized in table~\ref{models}, and are also shown on the plots of Fig.~\ref{H graphs}.

\section{Signatures from Sterile Neutrino Decays \label{Sec N Decays}}
%%%

Sterile neutrinos acquire couplings to the neutral and charged weak currents via the see-saw mechanism, and can thus decay into SM particles. For light sterile neutrinos, such as the keV dark-matter sterile neutrino, the only accessible channels are $N \rightarrow 3\nu$ and $N \rightarrow \gamma \nu$. Heavier sterile neutrinos, though, possess a variety of decay channels. While for masses less than 3~GeV, one has to consider decays into mesons~\cite{Gorbunov:2007ak}, here we focus on sterile-neutrino masses $\MN \gtrsim 3 \GeV$. Various aspects of the sterile-neutrino decay signals in colliders have been discussed in Refs.~\cite{Gorbunov:2007ak,Graesser:2007yj,Graesser:2007pc,Aranda:2007dq,Perez:2009mu,Han:2006ip,Bezrukov:2009th}. Here we summarize these signatures, with emphasis on sterile neutrinos with masses $\: {\rm few} \GeV \lesssim \MN \lesssim {\rm few} \times 100 \GeV$, and mixing with the active neutrino species $\theta^2 \lesssim 10^{-10}$. Such sterile neutrinos are allowed by the current astrophysical bounds and can contribute significantly to the observed neutrino masses~\cite{Smirnov:2006bu}. They can be produced in colliders if they couple to the Higgs sector, as already discussed, and may yield observable signals when they decay, as we will now see.

\begin{figure}[b]
\centering
\begin{tabular}{cc}
\epsfig{file=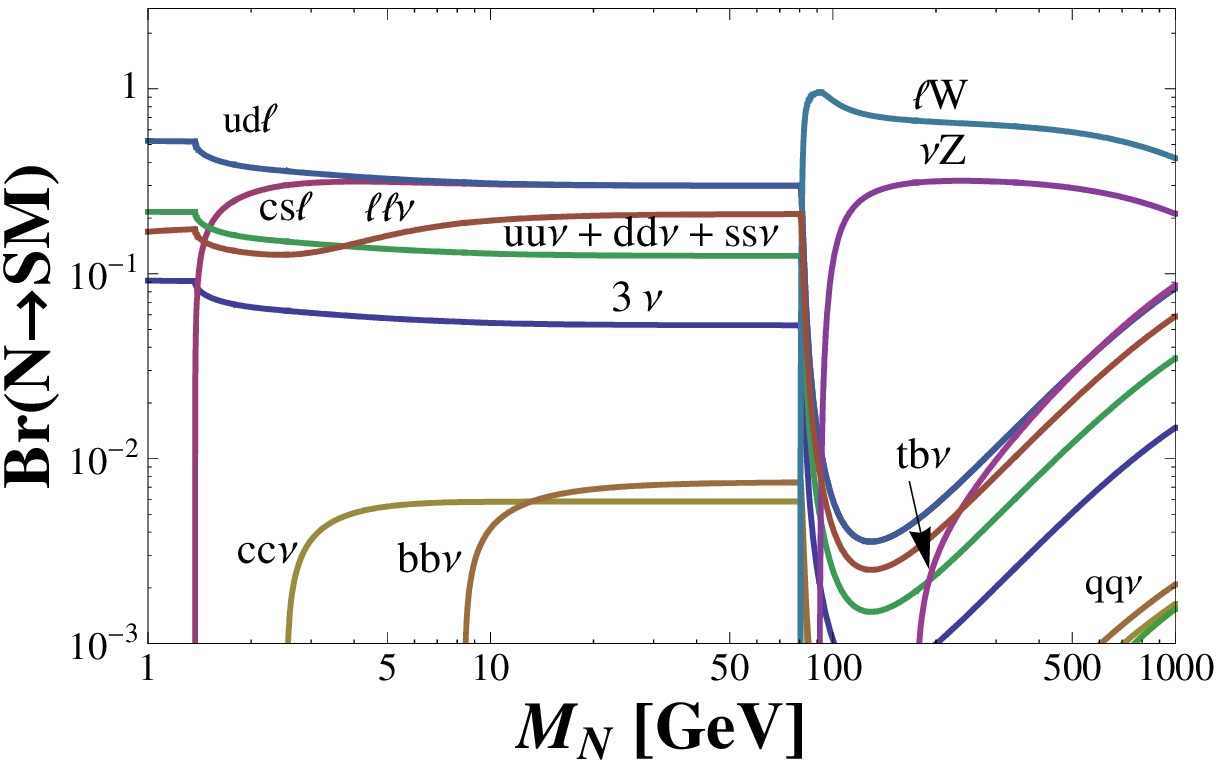,width=0.5\linewidth,clip=}
\epsfig{file=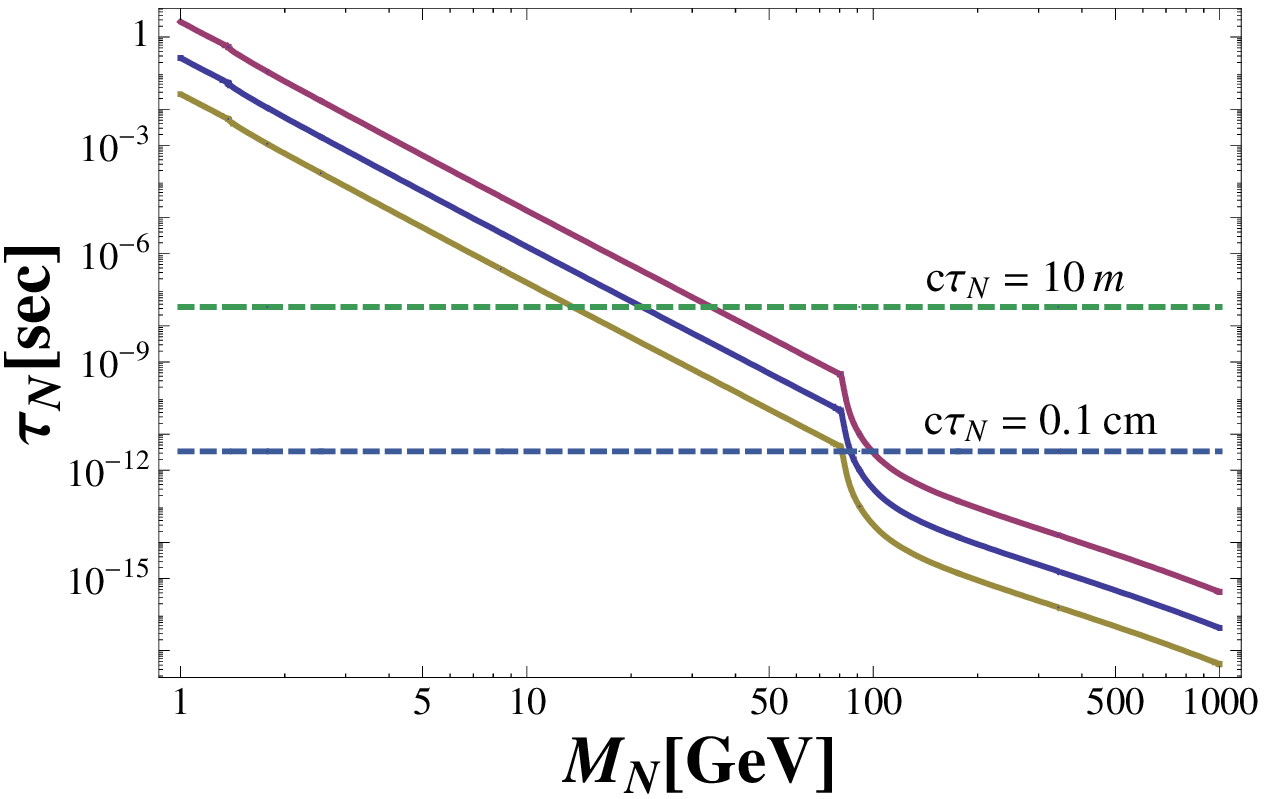,width=0.5\linewidth,clip=} \\
\end{tabular}
\caption{
\emph{Left}: Decay branching ratios of a sterile neutrino $N$ into the various SM channels (under the assumption that the $N \rightarrow \nu H_1 $ decay is kinematically forbidden). The graph is independent of the active-sterile neutrino mixing angle $\theta$.
\emph{Right}: Sterile-neutrino lifetime $\tN$ vs its mass. The solid lines correspond to different active-sterile mixing angles: $\theta^2 = 10^{-12}$ (purple, top line), $\theta^2 = 10^{-11}$ (blue, middle line), $\theta^2 = 10^{-10}$ (yellow, bottom line).
The kinks in the decay time occur when the decay into weak gauge bosons becomes kinematically accessible.
The horizontal dashed lines denote fixed values of the sterile-neutrino decay length. For  $c\tN \lesssim 0.1 \cm$, sterile neutrinos appear to decay promptly at the point of production; for $0.1 \cm \lesssim c\tN \lesssim 10 \mt$, their decays may give rise to displaced vertices; for $c\tN \gtrsim 10 \mt$, sterile neutrinos escape from the detector before they decay, and their production via Higgs decays will appear as missing energy.
}
\label{N graphs}
\end{figure}

The dominant decay mode of sterile neutrinos with mass $3 \GeV \lesssim \MN \lesssim 80 \GeV$, is into light quarks and a charged lepton, $N \rightarrow \ell_{\alpha}^{-} u\bar{d}$ and $N \rightarrow \ell_{\alpha}^{-} c\bar{s}$ and the charge-conjugated modes~\cite{Gorbunov:2007ak,Bezrukov:2009th}. Decay channels which are mediated by a virtual $Z$ boson are somewhat suppressed. These channels involve production of at least one lighter neutrino species which will appear as missing energy.
Above the $W$ production threshold, $\MN \gtrsim 80 \GeV$, the mode $N \rightarrow W^{\pm} \ell^{\mp}$ becomes dominant, and is shortly followed by the $N \rightarrow Z \nu$ mode.
We plot the decay branching ratios and the lifetime of sterile neutrinos in Fig.~\ref{N graphs}, using the decay rates provided in Ref.~\cite{Bezrukov:2009th}.

\bigskip

The sterile-neutrino lifetime is very important in determining the optimal search strategy.
It is given by~\cite{Bezrukov:2009th}
\be
\tN^{-1} \simeq  \left\{
\begin{array}{lr}
\sin^2 \theta \: \frac{G_F^2 \MN^5}{192 \pi^3} \: \alpha (\MN), & \quad \MN < M_{_{\rm W}}    \\
\sin^2 \theta \: \frac{G_F \MN^3}{4 \sqrt{2} \: \pi},           & \quad \MN > M_{_{\rm W}}
\end{array}
\right.
\label{Gamma N}
\ee
where $\alpha(\MN)$ accounts for the decay channels available for a sterile neutrino of mass $\MN$ (including the various $\mathcal{O}(0.1-1)$ factors associated with each of these channels). For $3 \GeV \lesssim \MN \lesssim 80 \GeV$, $\alpha \simeq 21.8$.
As the decay length $c \tN$ is very sensitive to the sterile-neutrino mass, we may discern the following cases:
\begin{enumerate}
\renewcommand{\theenumi}{\roman{enumi}}
  \item Sterile neutrinos with $\MN \lesssim  21 \GeV \left(\frac{10^{-11}}{\sin^2 \theta}\right)^{1/5}$ survive long enough to escape the detector before they decay, $c \tN \gtrsim 10 \mt$. Their production via Higgs decays will appear as missing transverse energy.
  \item If $\: 21 \GeV \left(\frac{10^{-11}}{\sin^2 \theta}\right)^{1/5} \lesssim \MN \lesssim 86 \GeV$,
%$\: 21 \GeV \left(\frac{10^{-11}}{\sin^2 \theta}\right)^{1/5} \lesssim \MN \lesssim 80 \GeV \left(\frac{10^{-11}}{\sin^2 \theta}\right)^{1/3}$,
  sterile neutrinos may decay inside the detector, but at distances visibly displaced from the production point, $0.1 \cm \lesssim c \tN \lesssim 10 \mt$.
  \item Heavier sterile neutrinos with masses $\MN \gtrsim  86 \GeV$
%$\MN \gtrsim  80 \GeV \left(\frac{10^{-11}}{\sin^2 \theta}\right)^{1/3}$
  decay promptly, at distances $c\tN \lesssim 0.1 \cm$ from the point of production.
\end{enumerate}

In Fig.~\ref{N graphs} we present the proper lifetime $\tN$ as a function of $N$ mass.  In the lab frame, the decay time and length will be of course dilated by the sterile-neutrino Lorentz factor.

The prospects of detecting an invisibly decaying Higgs have been assessed in Ref.~\cite{Eboli:2000ze}.
They find that even if the Higgs decays into invisible states with branching fraction of $\mathcal{O}(1)$, its production via Weak Boson Fusion ($qq\rightarrow qqVV \rightarrow qqH$) can be discovered at the $5\sigma$ level by imposing appropriate cuts on the correlation of the jet momenta.

A displaced vertex is a very indicative signature of a long-lived state. The kinematics of the decay products reconstruct to the location of the displaced vertex.
Displaced vertices emanating from sterile-neutrino decays of masses $\MN > M_W$ are dominated by the two-body decays with ${\rm Br}(N \rightarrow W^{\pm} \ell^{\mp}) \approx 88 \%$, and ${\rm Br}(N \rightarrow Z  \nu) \approx 12 \%$.
In Ref.~\cite{Graesser:2007pc} it was found that when the Higgs has $\mathcal{O}(1)$ couplings to a heavy sterile neutrino, displaced vertices smaller than $\sim 5~\cm$ are excluded by existing Tevatron bounds on the decay length. This is evaded, though, if the Higgs decays into sterile neutrinos with branching fraction around $0.1-0.01$. The limit is thus applicable mostly to a singlet-like Higgs state, which has large coupling and decay branching ratio to sterile neutrinos (such as $H_1$ in model C). Maximally-mixed or doublet-like Higgs states (e.g. $H_1$ in models A and B) comfortably avoid this constraint, and allow for $c \tau \lesssim 5 \cm$.

Prompt decays of sterile neutrinos yield an odd number of leptons and quark--anti-quark pairs. Due to their Majorana nature, the two sterile neutrinos produced in a Higgs decay can subsequently decay into a pair of leptons or a pair of antileptons, plus quark matter. This possibility is perhaps the most straightforward to discover by standard techniques since no vertex reconstruction is required, and each Higgs decay ends with a very dramatic signature of overall lepton number violation with branching ratios shown in Fig.~\ref{N graphs}. Since decays via charged currents dominate over decays via the neutral current, the most probable case of lepton-number violation is when the two sterile neutrinos decay into a pair of same-sign charged leptons plus jets.

%%%
\section{Conclusions \label{Sec concl}}
%%%

Neutrino masses can be incorporated to the SM by introducing new gauge-singlet fermions, the sterile neutrinos. These particles are associated with a new mass scale, which may arise via the Higgs mechanism, similarly to the fermion masses in the SM. If the symmetry breaking associated with sterile-neutrino masses occurs also at the electroweak scale, there can be a number of interesting implications both for cosmology and particle physics at colliders.

In its simplest implementation, the above scenario suggests that sterile neutrinos couple to a real gauge-singlet scalar field, which is part of the Higgs sector and picks up a VEV after EWSB. Light sterile neutrinos, of mass of a few keV, will be weakly coupled to the singlet Higgs, and may be produced via decays of the latter in the correct abundance to account for the DM of the universe. Heavy sterile neutrinos will couple strongly to the Higgs sector, and will yield interesting signatures for upcoming accelerator experiments.

In this paper, we outlined the collider signatures that arise from this simple extension of the SM. Sterile neutrinos responsible for the observed neutrino masses are typically anticipated to have very weak mixing with the active neutrino sector, which suppresses the prospects for their detection. However, their coupling to the Higgs sector renders such particles discoverable at colliders. In this case, sterile neutrinos may produce a number of very suggestive signals, such as displaced vertices, invisible decays, and lepton-number violating decays with production of same-sign charged leptons. Detailed simulations are, of course, needed. The detection of sterile-neutrino signatures at colliders would then imply the existence of an extended Higgs sector.

The coupling of the SM Higgs with a singlet scalar relaxes the bounds on the Higgs mass, and thus alleviates the tension between the current LEP and EWPO limits. This effect is enhanced by the coupling of sterile neutrinos to the Higgs sector. A potential discovery of a Higgs boson with mass outside the SM-allowed region would point towards new couplings of the Higgs to non-SM particles, possibly extra singlet scalars and/or singlet fermions.

\section*{Acknowledgments}
We would like to thank Michael Graesser for reading a draft of the manuscript and for very helpful comments and suggestions.  This work was supported in part by DOE grant DE-FG03-91ER40662 and NASA ATP grant NNX08AL48G. KP was supported, in part, by the Australian Research Council.

\bibliographystyle{JHEP}
\bibliography{neutrino}

\end{document}